\documentclass[12pt]{article}
\usepackage{graphicx}
\usepackage{epsfig}

\usepackage{amssymb}
\begin{document}

\title{The Local Fractal Properties of the Financial Time Series on the Polish Stock Exchange Market}
\author{Dariusz Grech\footnote{dgrech@ift.uni.wroc.pl} \  and \  Grzegorz Pamu\l a \\
Institute of Theoretical Physics\\
University of Wroc{\l}aw, PL-50-204 Wroc{\l}aw, Poland}

\date{}
\maketitle

\begin{abstract}
We investigate the local fractal properties of the financial time
series based on the evolution of the Warsaw Stock Exchange Index
(WIG) connected with the largest developing financial market in
Europe. Calculating the local Hurst exponent for the WIG time series
we find an interesting dependence between the behavior of the local
fractal properties of the WIG time series and the crashes appearance
on the financial market.
\end{abstract}

The models and mechanisms enabling to predict the future behavior of
the financial market on a long or short term period are a big
challenge in the financial engineering and very recently also in
econophysics. In the latter case one believes that the approach
based on the analogy of the financial market with the complex
dynamical system could be very fruitful \cite{grepa1}--\cite{grepa8}. In
particular, the scale invariance of the complex systems is used to
reveal the log-periodic oscillations characteristic for these
systems before the phase transition point is reached. The
log-periodic oscillations preceding the crashes or ruptures points,
\emph{i.e.} the moments when the increasing long-term trend is being
broken and starts to reverse, have really been observed for the
market indices or share prices (see
\emph{e.g.} \cite{grepa1,grepa2}, \cite{grepa9}--\cite{grepa13}). A number of crash
moments $t_c$ has been predicted in the literature so far, some of
them being in a very good agreement with the actual moment the crash
took place \cite{grepa11}--\cite{grepa13}.\\
However one has to remember that any financial system should be
considered as an open system, while the scaleless behavior and the
quantitative description following from this assumption are
completely true only in a case of the closed statistical systems.
Therefore one should be aware that the methods based on the closed
system assumption are very sensitive to the number of data points
(information) one takes into account to make the fit of log-periodic
oscillation parameters. As a result the predictive power of such
models can be in general very limited \cite{grepa14}.

Few years ago another approach based on the local properties of the
time series was proposed \cite{grepa15}. The method uses the local
Hurst exponent $H_{loc}$ \cite{grepa16,grepa17} or the local fractal
dimension $D_{loc}$ of the time series built on the index values or
share prices. These quantities are linked together by the well known
relation
\begin{equation}
D_{loc} = 2 - H_{loc}
\end{equation}
in an analogy with the similar equation satisfied for the global $D$
and
$H$ values usually used to describe the monofractal signals.\\
 In Ref.~\cite{grepa15} it was shown that the local Hurst exponent calculated
 for the main Dow Jones (DJIA) index forms the characteristic pattern
 before any of the crashes on American stock market.
 The  $H_{loc}$ values drop significantly down before any rupture
 point. The moving average $\langle H_{loc}(t)\rangle_5$ of the local
 Hurst exponent calculated on the one week period (5 sessions) drops
 down to $0.45$ or even less for sessions immediately preceding the
 rupture point, thus revealing the presence of the growing
 antipersistence in the financial time series signal before
 the crash occurs. The similar qualitative behavior of the local $H$
 values for other financial time series (shares) has also been
 confirmed \cite{grepa18}.\\
 The advantage in the use of $H_{loc}$ over other methods is that it
 actually measures 'the local state' of the market and therefore it
 seems to be more resistable to the long-term
 inaccuracies or distortions coming \emph{e.g.} from the rapid change
 of the boundary conditions around the financial system.
 Therefore the ${H_{loc}}$ method might be also applied to an open complex system
 and the financial
 market is a good candidate of such system.\\
 In this report we intend to apply the technique used in Ref.~\cite{grepa15} to
 investigate the market index of the largest emerging market in
 Europe - the Warsaw Stock Exchange Index (WIG)
 incorporating more than $200$ companies. It has already $15$-years
 old history up to now with more than $3600$ closure day values at
 the moment.\\
 The WIG time series has been analyzed by us with the
 Detrended Fluctuation Analysis (DFA) technique \cite{grepa16} to extract the
 scaling Hurst exponent $H$. This technique is well described in the
 literature \cite{grepa16,grepa17,grepa19} so we will not quote it here. Let us remind however
 its local version applied in our calculations.\\
 We first form for a given trading day $t=i$ a time subseries of
 length $N$ with points in the period $\langle i-N+1, i\rangle$. We call this subseries
 the observation box or the observation time-window. Then the
 standard DFA procedure is applied to this time-window, \emph{i.e.} we
 cover the subseries with smaller non-overlapping  boxes of size $\tau$
 starting from the given trading point $t=i$ and going backwards in
 time up to $t=i-N+1$. In order to cover the whole time-window with $\tau$-size boxes
 we put the last box in the period $\langle i-N+1, i-[N/\tau]\tau +1\rangle$,
 where $[.]$ means the integer part. This box partly overlaps the
 preceding one but this does not modify the obtained results. In
 each box the detrended signal is found according to DFA method for
 the simplest linear trend assumed in every box of size $\tau$. The
 detrended signal fluctuates and its variance $ \langle F^2(\tau)\rangle$ is related to the box
 width $\tau$ by the power law relation known for DFA:
 \begin{equation}
 \langle F^2(\tau)\rangle \sim \tau^{2H_{loc}}
 \end{equation}
Thus moving the observation box of length $N$ session after session
we are able to reproduce the whole history of $H_{loc}(t)$ changes
in
time.\\
It is well known that the $H$ exponent measures  the level of
persistence $(H>1/2)$ or anti persistence $(H<1/2)$ in the signal.
For $H=1/2$ one obtains the Brownian (integer) signal with null
autocorrelations. Hence, only the case $H\neq 1/2$ is meaningful for
practical applications. The observation of the $H_{loc}(t)$
evolution may suggest in what state the financial market is at the
given moment. Indeed, big investors usually called speculators cash
their profits more frequently if they 'feel' the rupture point in
the increasing trend is coming. The more nervous behavior of
speculators gives a signal to all other players on the market
(spectators) who also start to cash their investments at lower
prices. It actually leads after some time to the change of trend and
if the market is particularly nervous - to the crash appearance.
Thus we put a hypothesis that the nervous situation on the market
can (should) be observed as the appearance of the anti correlations
in price returns of various assets and finally we may expect the
growing local fractal dimension of the financial time series (or
decreasing
$H_{loc}$) before a crash.\\
To check if this hypothesis works well for the polish financial
market we studied three main crashes (or rupture points) that have
already taken place during the whole WIG history 1992-2007. First we
calculated the $H_{loc}$ for any point of the closure day WIG time
series. The $H_{loc}$ are usually widely spread out due to the
statistical noise. Therefore the moving average $\langle
H_{loc}\rangle_5$ of last five sessions (one trading week) or the
moving average $\langle H_{loc}\rangle_{21}$ of last 21 sessions
(one trading month) have been calculated for the local Hurst
exponent for the whole WIG history. The results of $\langle
H_{loc}\rangle_{21}$ are shown in Fig.~1 and Fig.~2. All plots were
done for the observation boxes of length from $N=215$ till $N=300$
sessions. It corresponds to the depth of looking back in time from
10 months till 15 months.\\
Let us notice that the main trend pattern of $H_{loc}(t)$ is
independent on the observation box length $N$ but the depth of
$H_{loc}$ fluctuations does depend on $N$ (see Fig.~2). It is
because of the different noise cut level changing with the
time-window length. To examine the structure of a local behavior
more rigorously we have chosen the 10 months observation box to
perform further analysis (the similar size was used in Ref.~\cite{grepa15}). The
examples of plots $F^2(\tau)$ vs $\tau$ in log-log scale, from which
the $H_{loc}$ values have been extracted
according to Eq.~(2), are shown in Fig.~3.
\begin{table*}[h]
\caption{
The main crashes on the Polish financial market. The total
percentage drop in WIG and its duration time is shown as well as the
percentage drop in the first three sessions after the rupture
point.
}
\begin{tabular}{||l|l|l||}
\hline
Date &Initial 3 sessions drop &Total relative drop(duration
time)
\cr   \hline
17.03.94 & 11 \% & 65 \% (41 sessions)
\cr\hline
22.07.98 & 4 \%       & 39 \%      (30 sessions)
\cr\hline
12.05.06 & 5 \%          & 21\%      (24 sessions)
\cr
\hline
\end{tabular}
\end{table*}

Then the most spectacular crashes on the Polish stock exchange
market were examined within this technique in some surrounding of
the crash (rupture) points. The investigated cases are found between
the vertical lines in Fig.~1. and correspond to crashes described in
Table 1. The zoomed evolution of the $H_{loc}(t)$ for these crashes
is shown in Fig.~4a-c. The decreasing trend in $H_{loc}$ is the most
evident from these plots and lasts for many sessions before the
rupture point occurs. Looking at the common characteristic pattern
of $H_{loc}$ plots before the crash moment, we may formulate the
following necessary conditions to be \emph{simultaneously} satisfied
(\emph{signal to sell}) if the rupture point is expected soon:
\begin{enumerate}
\item{} $H_{loc}(t)$ is in decreasing trend and $\langle
H_{loc}\rangle_5 <\langle
 H_{loc}\rangle_{21}$
  except for small fluctuations
\item{} $\langle
H_{loc} \rangle_{21}\leq 0.5$
\item{} $\langle H_{loc}\rangle_5
 \lesssim  
0.45$
\item{} minima of $H_{loc}(t)$
 for not necessary consecutive sessions satisfy
 $H_{loc}^{min}(t) \lesssim 0.4$
\end{enumerate}
Contrary, if all the above conditions are not satisfied we expect
the strong \emph{signal to buy} on the market.\\
Moreover we are able to find a relation between the rate of the
$\langle H_{loc}\rangle_5$ drop and the total correction the WIG
index gains after the crash. Generally, the deeper the bottom of $H_{loc}(t)$ signal
the bigger crash or major correction can be expected. The latter correction can be calculated
as the difference in the index signal magnitude between the rupture point
and the minimum in the signal value after the crash from which the
next long-lasting trend is being formed. We observe this correction
is proportional to the absolute slope of the $H_{loc}$ trend assumed
as a straight line. Such straight line-fits to the local Hurst
exponents have been drawn in Fig.~4a-c for periods immediately
preceding the crashes on the Polish stock exchange market. We have
also drawn in Fig.~5 a relation between the magnitude of the relative
WIG index correction and the slope of the $H_{loc}(t)$ linear trend
fit before the crash. Amazingly, this relation is linear! The fit done with just three points
may not of course entitle to draw a strong conclusion,
nevertheless the relation is striking and worth further
investigation.\\
Finally we checked also the current situation on the market. The
plot of the recent WIG signal with the corresponding local fractal
properties is illustrated in Fig.~6. The $H_{loc}(t)$ started to fall
down almost three trading months ago (from the session $\sharp
3561$), despite the WIG signal has still been increasing. The
$\langle H_{loc}\rangle_5$ average reached its critical value $1/2$
in the end of June (June 28 '07 -- session $\sharp 3600$), however the
crash pattern (conditions (1)-(4)) is not so clear as in previous cases (a)--(c).
We still have $\langle H_{loc}\rangle_{21} \sim
1/2$ remaining only slightly below the critical value $1/2$. Also $H_{loc}(t)$ minima
are higher then previously. As a result a crash has not been
formed so far but the $7\%$ correction in WIG signal
actually took place. The future situation on the market will, in our
opinion, be driven by the forthcoming behavior of $H_{loc}(t)$ as
indicated in Fig.7. One may think about two possible scenarios. In
the first one the local Hurst exponent will remain in the decreasing
trend. If the dropping rate of $H_{loc}(t)$ remains the same as it
is now we may expect $\langle H_{loc}(t)\rangle \sim 0.4$ around the
beginning of September '07 (see Fig.7). This means that a major
crash should take place no later than mid-September. If however, we
would observe that $\langle H_{loc}\rangle$ began to increase and the other
conditions (1)--(4) were not fulfilled, the
crash should not take place and the current drop of WIG index will
be just a minor correction in the long-lasting increasing trend of
WIG signal. We believe the further $H_{loc}(t)$ evolution will choose very soon
one of these scenarios.

\newpage

\begin{figure}
\begin{center}
{\psfig{file=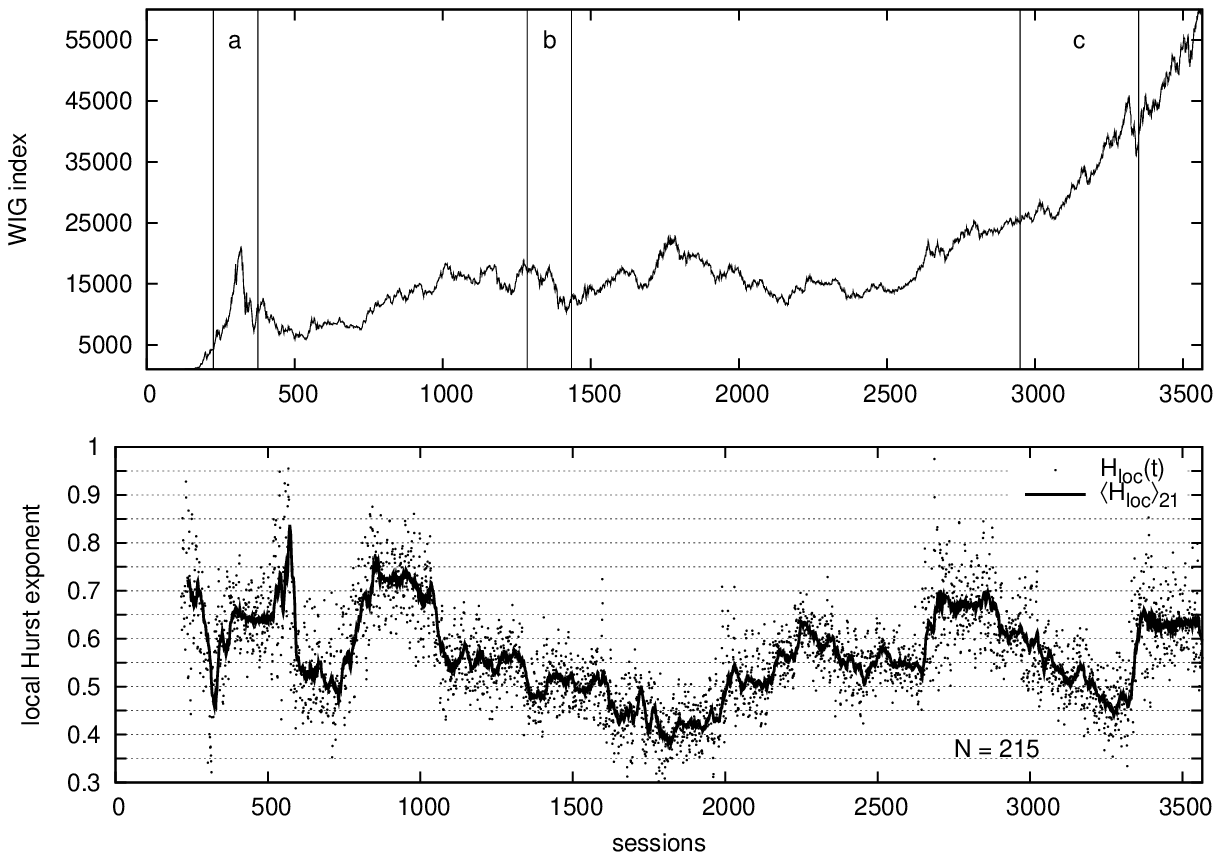,width=17cm,angle=90}}
\end{center}
\caption{The closure day WIG history April'92-May'07 (top) and the
corresponding local Hurst exponent (bottom). The time-dependent
Hurst exponent has been calculated in the observation box of $N=215$
sessions. The solid line represents the moving average of $H_{loc}$
(marked as dots) calculated for one trading month back ($21$
sessions). Three main crash periods are marked within the vertical
lines as (a), (b), (c). 
}
\end{figure}

\begin{figure}
\begin{center}
{\psfig{file=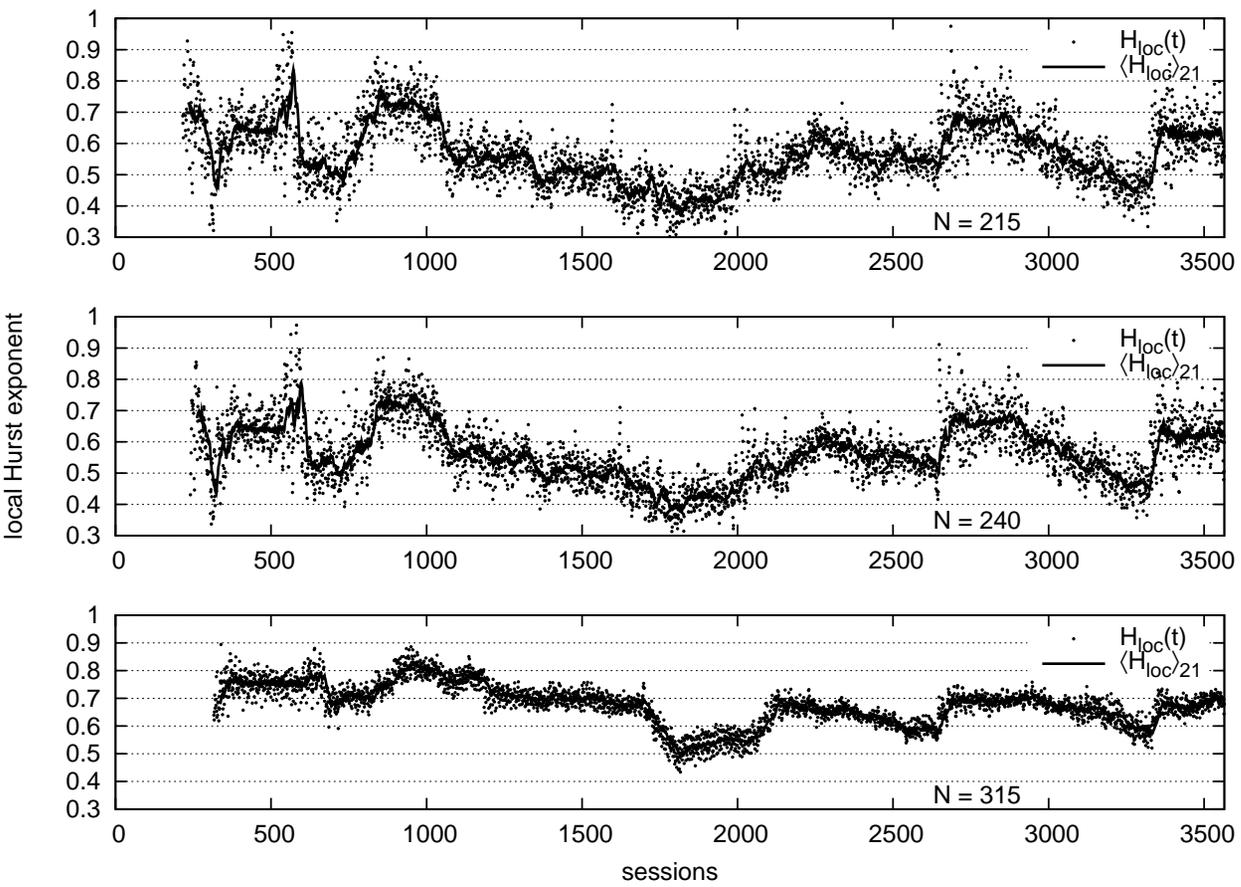,width=17cm,angle=90}}
\end{center}
\caption{Comparison
of the local Hurst exponents obtained for three different
observation boxes of length $N=215, N=240$ and $N=315$ sessions
respectively. The solid line represents the
$1$-month moving average of $H_{loc}(t)$ 
}
\end{figure}


\begin{figure}
\begin{center}
{\psfig{file=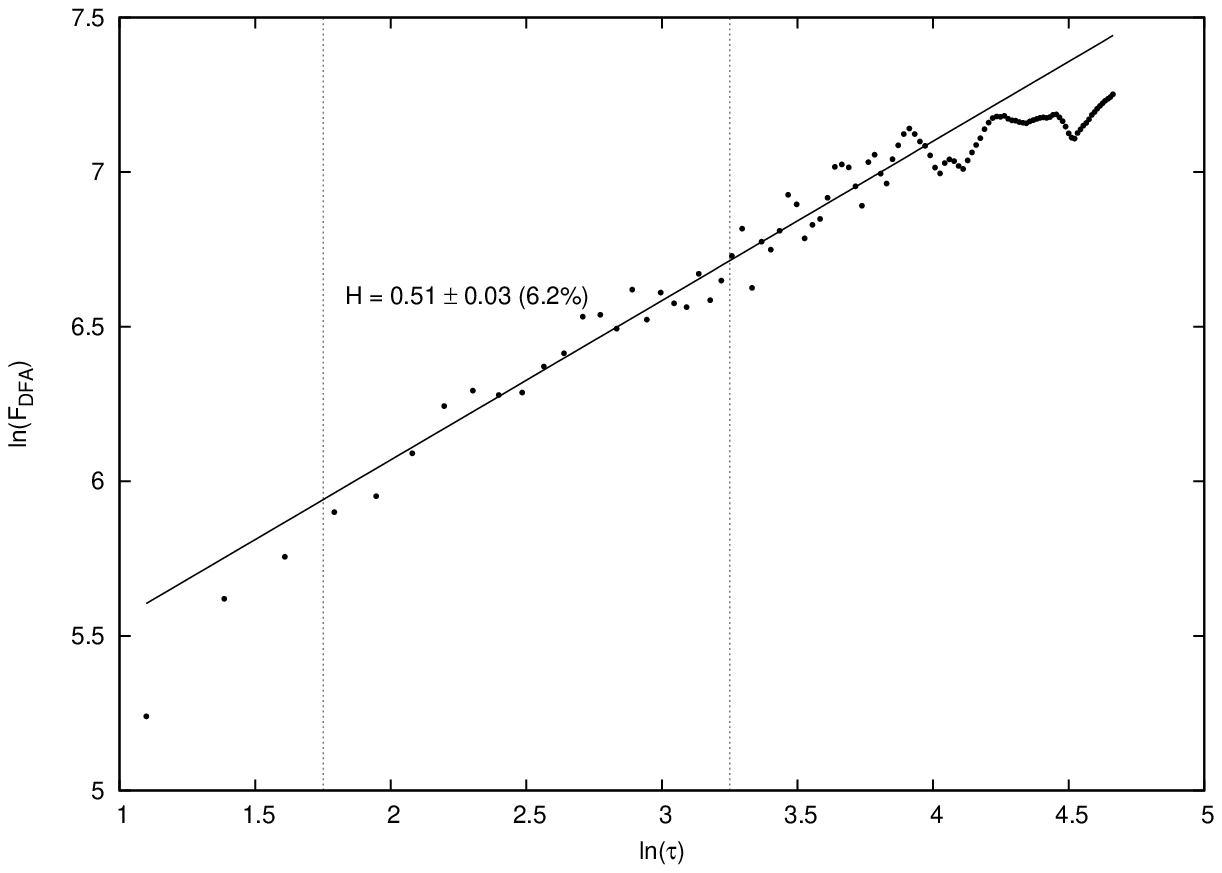,width=7.5cm,angle=0}}
{\psfig{file=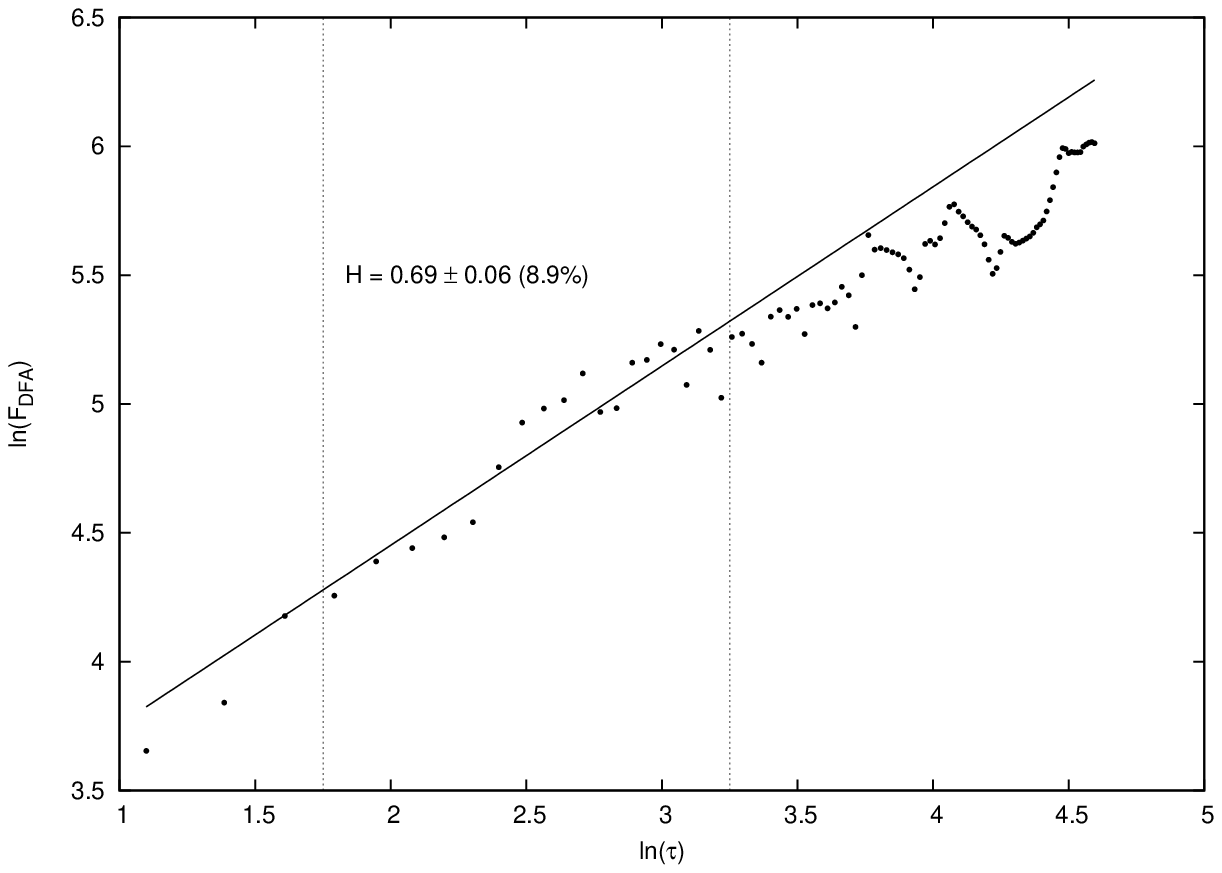,width=7.5cm,angle=0}}
\end{center}
\begin{center}
{\psfig{file=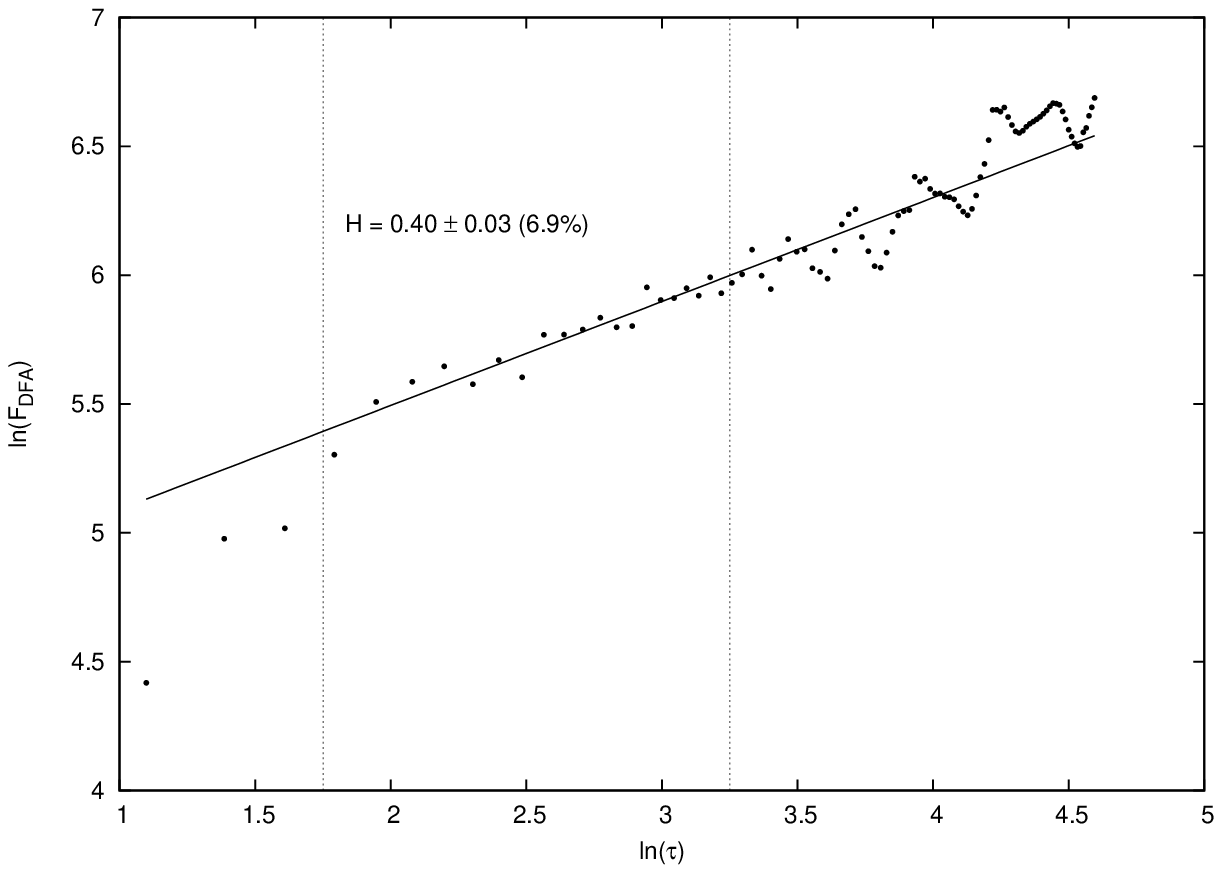,width=7.5cm,angle=0}}
{\psfig{file=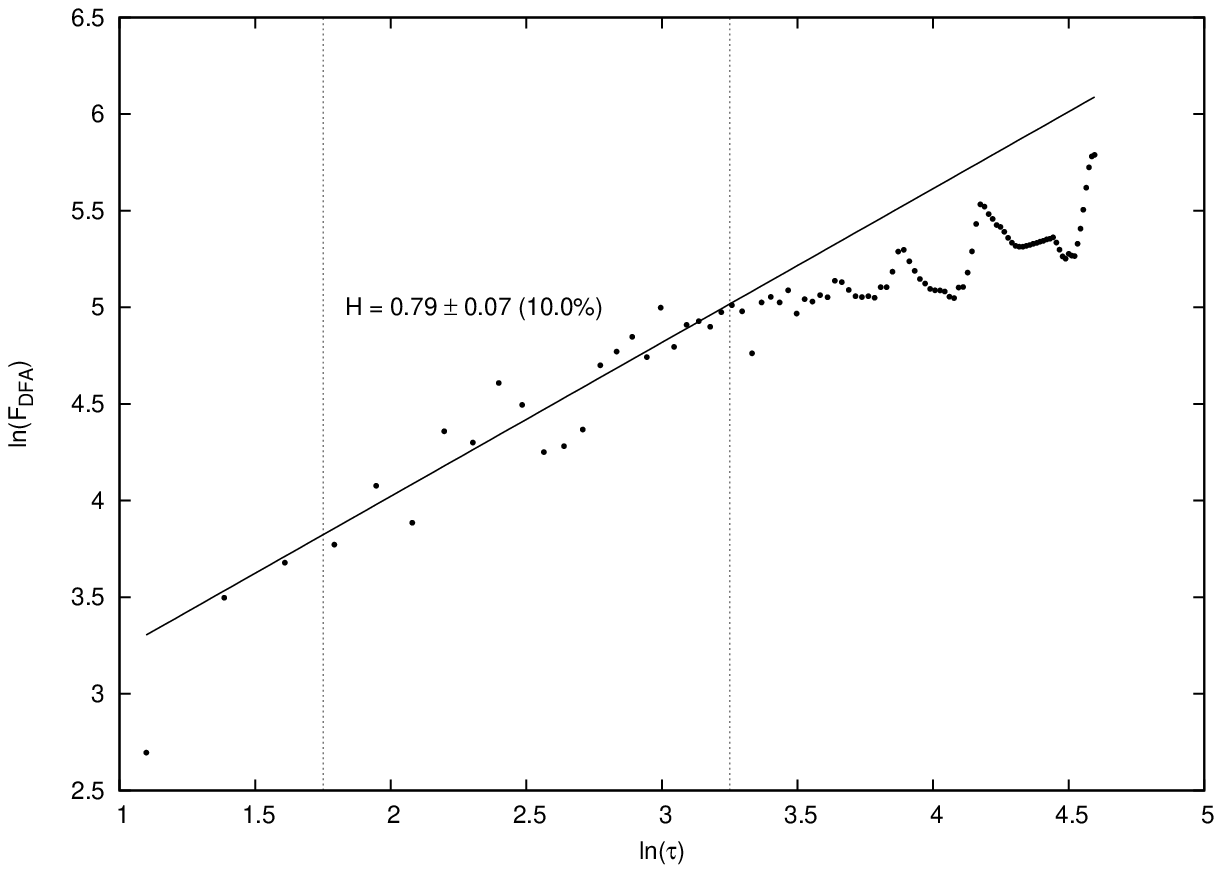,width=7.5cm,angle=0}}
\end{center}
\caption{Examples
of the local Hurst exponent calculated for the observation box of
length $N=215$ ($10$ trading months). Plots show the dependence
between fluctuation (variance) of the detrended WIG
signal in a box of size $\tau$ and the size of the box in a log-log scale. The dotted vertical lines mark
the scaling range for the power-law from
Eq.~(2).
}
\end{figure}
\renewcommand{\thefigure}{4\alph{figure}}
 \setcounter{figure}{0}
\begin{figure}
\begin{center}
{\psfig{file=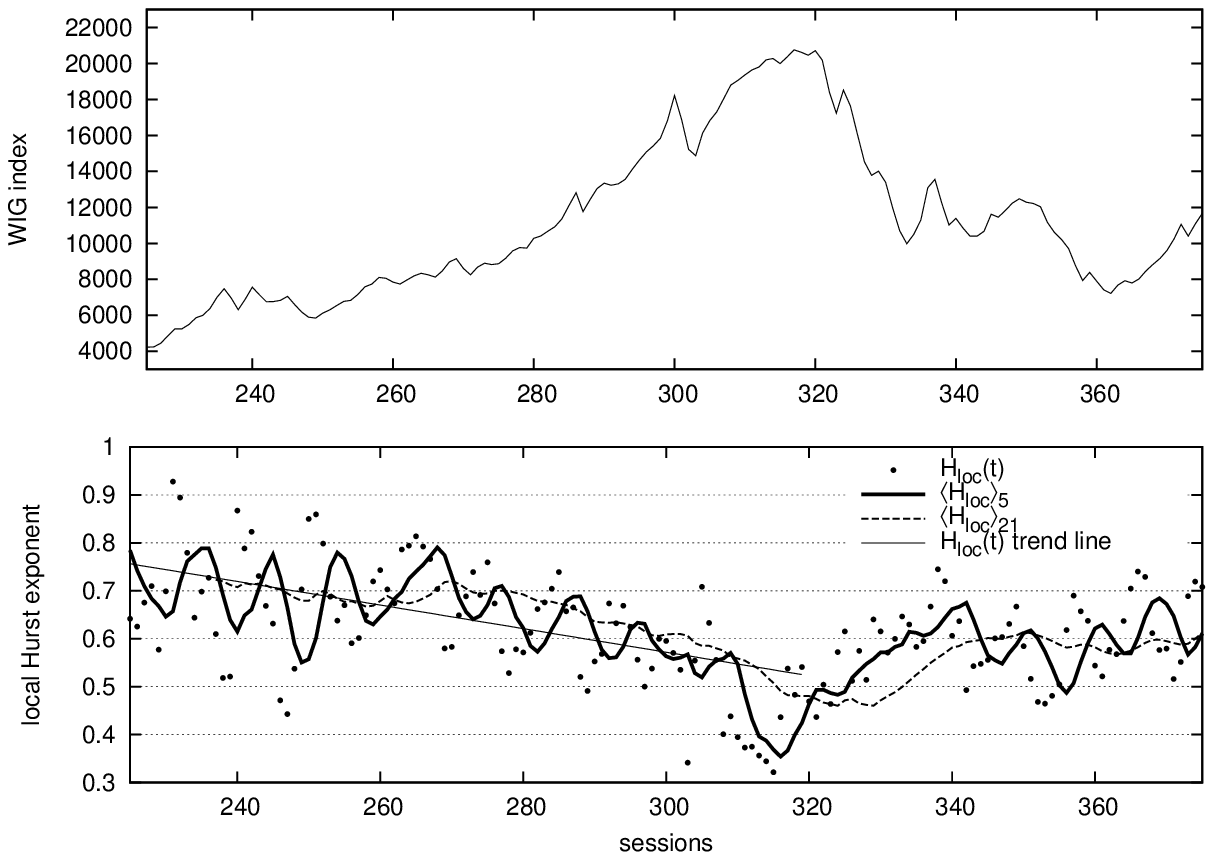,width=17cm,angle=90}}
\end{center}
\caption{
The
'X-ray' of major crashes (rupture points) on the Polish market done
with the local Hurst exponent -  March'94.
The dots represent $H_{loc}$ values, solid lines indicate
the one week $\langle H_{loc}\rangle_5$ and one month $\langle
H_{loc}\rangle_{21}$ moving averages of $H_{loc}(t)$. The line-fit
of the decreasing trend for $H_{loc}(t)$ before the crash is
also marked.
}
\end{figure}

\begin{figure}
\begin{center}
{\psfig{file=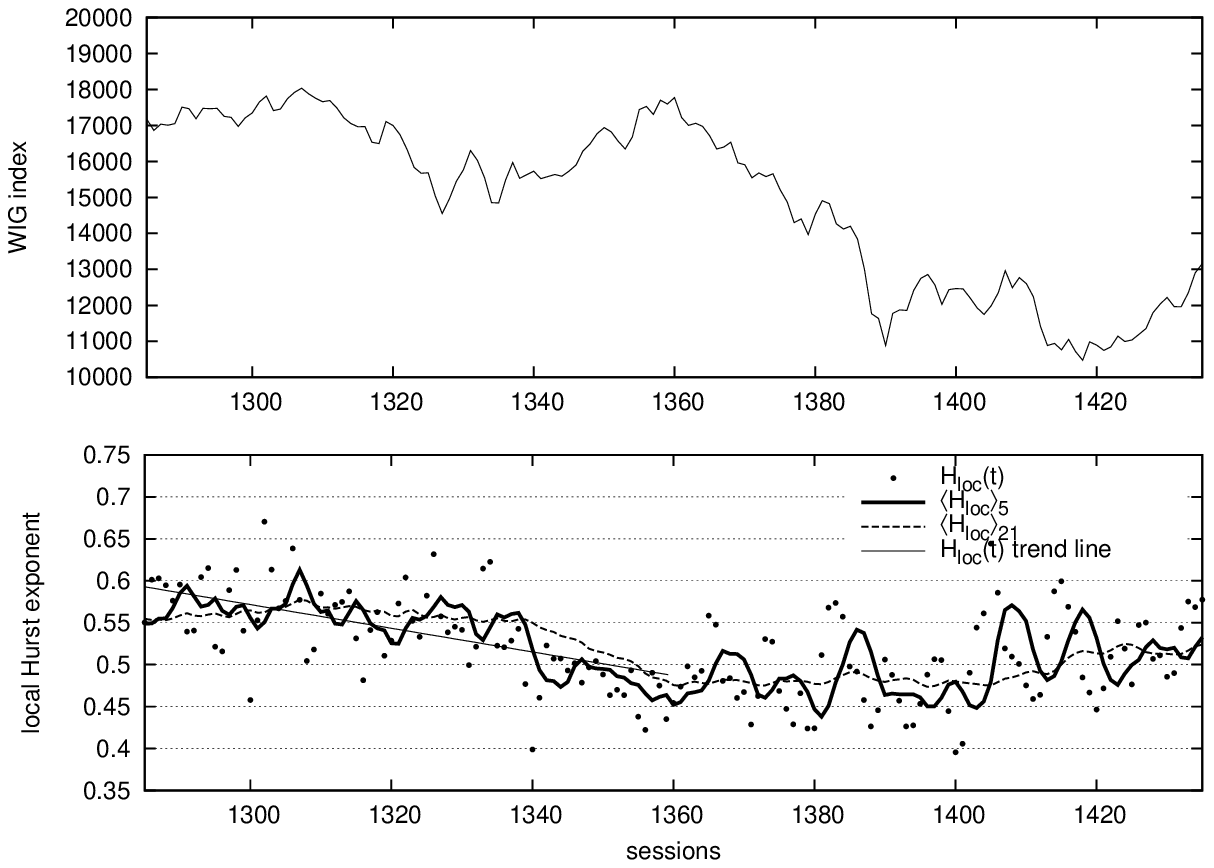,width=17cm,angle=90}}
\end{center}
\caption{
The
'X-ray' of major crashes (rupture points) on the Polish market done
with the local Hurst exponent - July'98.
The dots represent $H_{loc}$ values, solid lines indicate
the one week $\langle H_{loc}\rangle_5$ and one month $\langle
H_{loc}\rangle_{21}$ moving averages of $H_{loc}(t)$. The line-fit
of the decreasing trend for $H_{loc}(t)$ before the crash is
also marked.
}
\end{figure}

\begin{figure}
\begin{center}
{\psfig{file=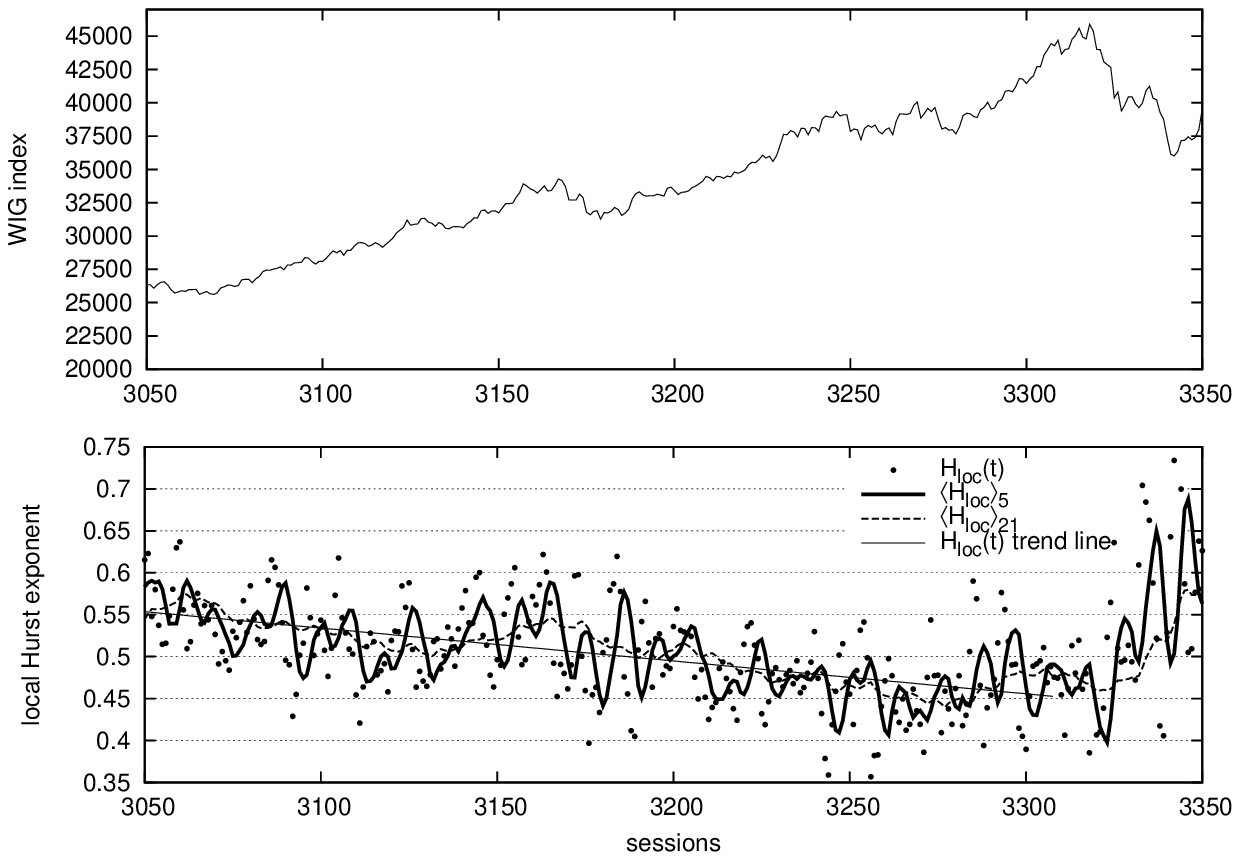,width=17cm,angle=90}}
\end{center}
\caption{
The
'X-ray' of major crashes (rupture points) on the Polish market done
with the local Hurst exponent -
 May'06.
The dots represent $H_{loc}$ values, solid lines indicate
the one week $\langle H_{loc}\rangle_5$ and one month $\langle
H_{loc}\rangle_{21}$ moving averages of $H_{loc}(t)$. The line-fit
of the decreasing trend for $H_{loc}(t)$ before the crash is
also marked.
}
\end{figure}
\renewcommand{\thefigure}{\arabic{figure}}
 \setcounter{figure}{4}
\begin{figure}
\begin{center}
{\psfig{file=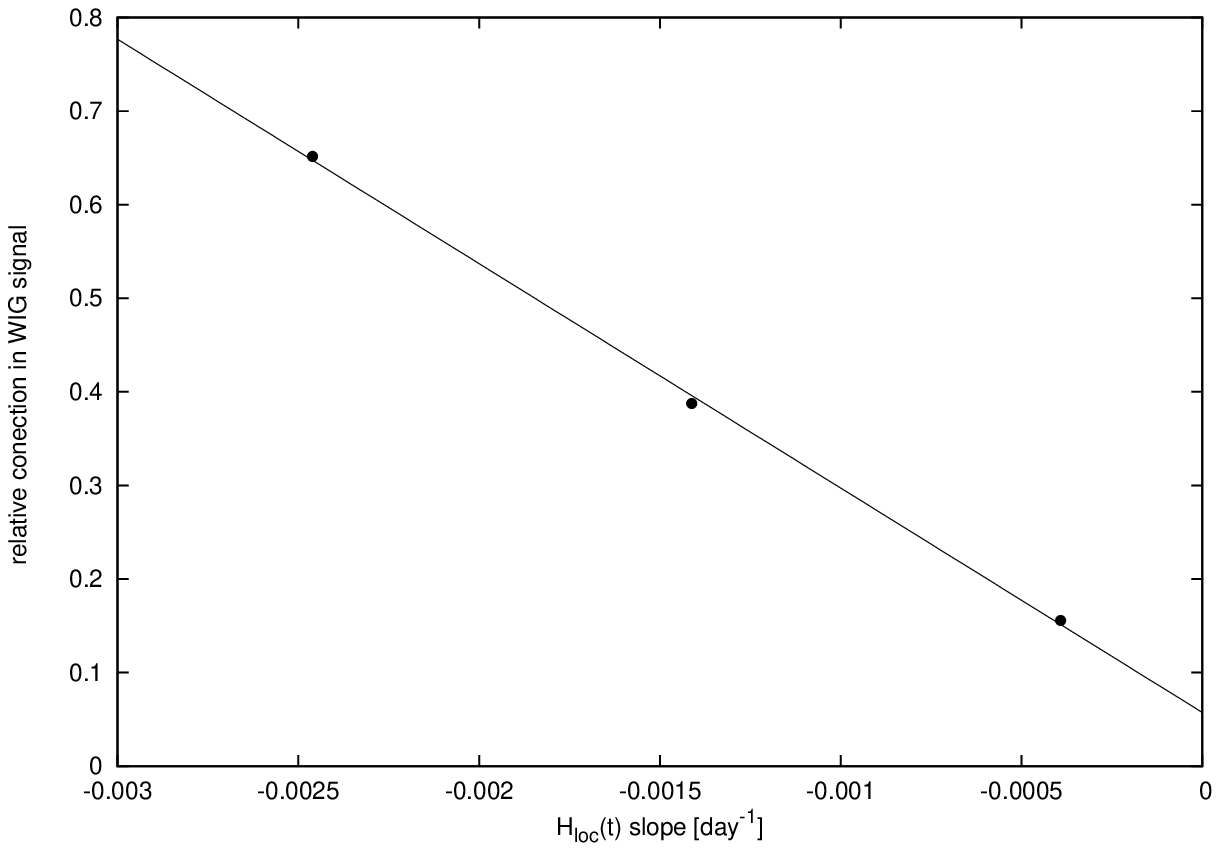,width=15cm,angle=0}}
\end{center}
\caption{The
dependence between the magnitude of correction in WIG signal
after the crash and the slope of line-fit to the $H_{loc}(t)$ trend.
}
\end{figure}
\begin{figure}
{\psfig{file=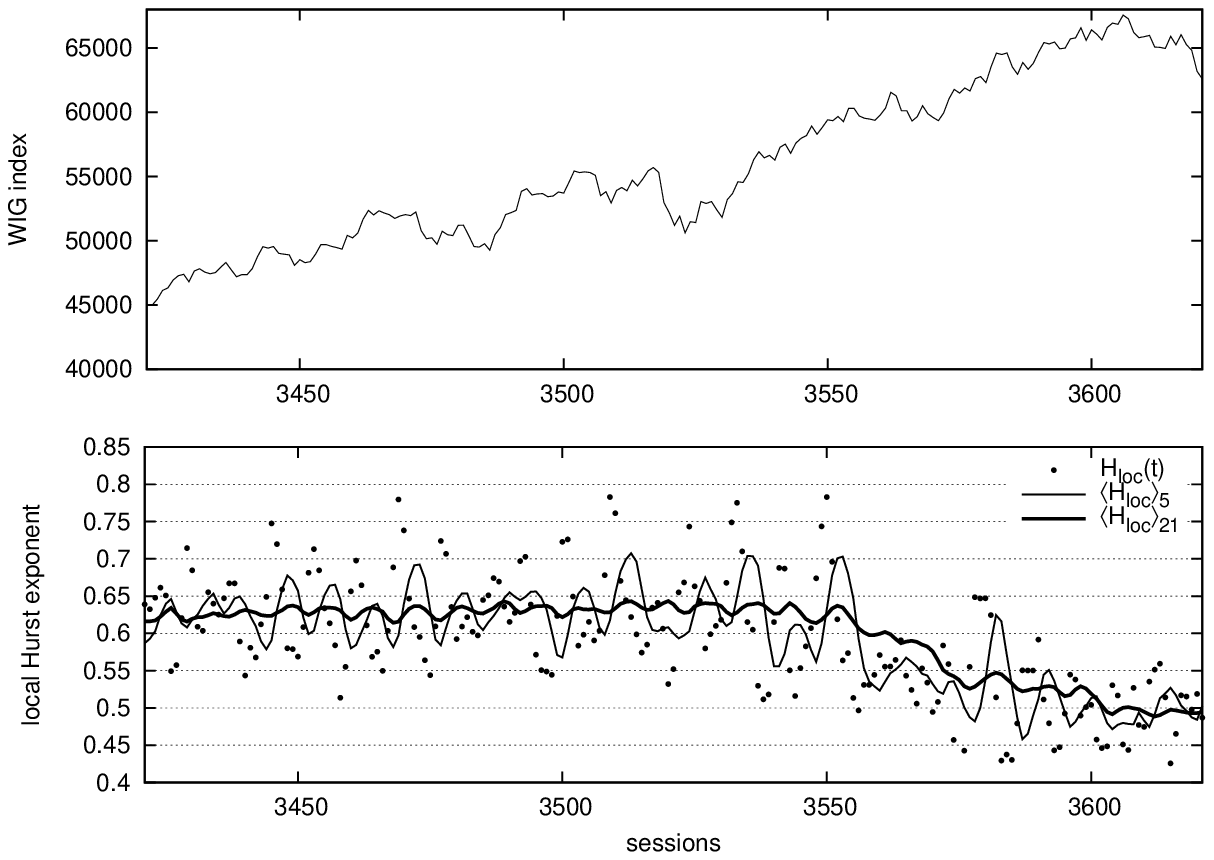,width=15cm,angle=0}} \caption{The recent
situation on the Polish market and the $H_{loc}(t)$ evolution.}
\end{figure}

\begin{figure}
{\psfig{file=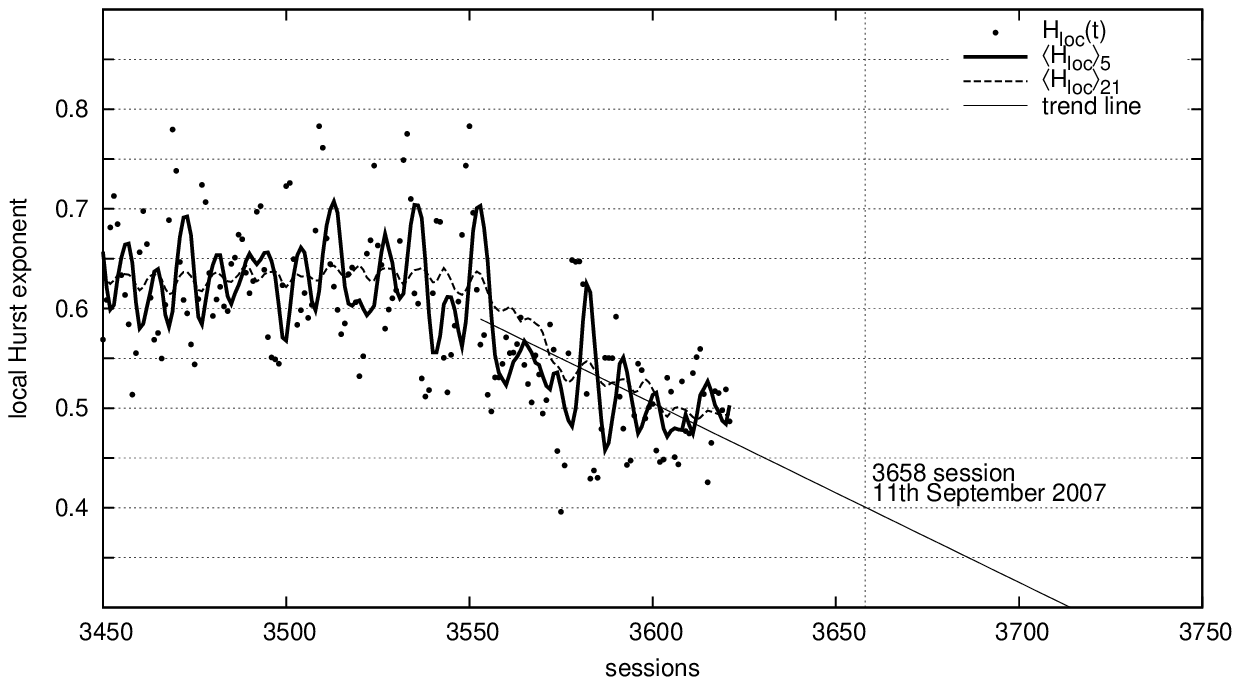,width=15cm,angle=0}}
 \caption{The possible
scenario of the further $H_{loc}(t)$ evolution if the decreasing
trend in local Hurst exponent will be kept. The data taken into
account terminate on July 27 '07. The straight line fit to
$H_{loc}(t)$ was done for $3$ trading months back.
}
\end{figure}
\end{document}